\documentclass[10pt,a4paper,twoside]{article}
\usepackage[english]{babel}
\usepackage[latin1]{inputenc}
\usepackage{amssymb}
\usepackage{mathrsfs}
\usepackage{amsmath}
\usepackage[dvips]{graphicx}
\usepackage{epsfig,bm}
\usepackage{graphicx}
\usepackage{rotate}
\usepackage{vmargin}

\usepackage{amsfonts}
\usepackage{graphics}
\usepackage{theorem}

\usepackage{color}
\usepackage{setspace}
\usepackage{amscd}
\usepackage{pstricks}
\usepackage{lscape}

\title{Coupled symplectic maps as models for subdiffusive processes in disordered Hamiltonian lattices}
\author{Chris G. Antonopoulos$^{1}$, Tassos Bountis$^{2}$, and Lambros Drossos$^{3}$}

\date{}

\begin{document}

\maketitle

\begin{center}
$^{1}$Institute for Complex Systems and Mathematical Biology (ICSMB), SUPA, Department of Physics, University of Aberdeen, AB24 3UE Aberdeen, United Kingdom\vspace{0.5cm}
\\
$^{2}$Center for Research and Applications of Nonlinear Systems, Department of Mathematics, University of Patras, 26500 Patras, Greece\vspace{0.5cm}
\\
$^{3}$High Performance Computing Systems Lab (HPCS lab), Department of Computer and Informatics Engineering, Technological Educational Institute of Western Greece, 30300 Antirion, Greece
\par\end{center}

\begin{abstract}
\noindent We investigate dynamically and statistically diffusive motion in a chain of linearly coupled 2-dimensional symplectic McMillan maps and find evidence of subdiffusion in weakly and strongly chaotic regimes when all maps of the chain possess a saddle point at the origin and the central map is initially excited. In the case of weak coupling, there is either absence of diffusion or subdiffusion with $q>1$-Gaussian probability distributions, characterizing weak chaos. However, for large enough coupling and already moderate number of maps, the system exhibits strongly  chaotic ($q\approx 1$) subdiffusive behavior, reminiscent of the subdiffusive energy spreading observed in a disordered Klein-Gordon Hamiltonian. Our results provide evidence that coupled symplectic maps can exhibit physical properties similar to those of disordered Hamiltonian systems, even though the local dynamics in the two cases is significantly different.
\end{abstract}


\vspace{0.5cm}
\noindent{\em In memory of our beloved friend and distinguished scientist Professor Theodore Papatheodorou, who taught us so eloquently that numerical analysis is indispensable when one tries to penetrate the mysteries of realistic physical systems.} 

\section{Introduction}\label{intro}

The absence of diffusion in disordered media (the so-called \emph{Anderson localization} \cite{Ander}) is an important effect that arises in transport phenomena, such as electromagnetic, acoustic and spin waves, in different types of classical or quantum linear systems. It is more interesting, however, to study what happens to the disordered system if nonlinearity is introduced. Understanding the effect of nonlinearity on the localization properties of wavepackets in disordered systems has attracted the attention of many researchers to date \cite{pikovsky_destruction_2008,kopidakis_absence_2008,Flach2009,Skokos2009,veksler_spreading_2009,SF10,Flach_spreading_2010,Laptyeva_crossover_2010,VKF10,MP10,MAP11,BLSKF11,BLGKSF11,BouSkobook2012}. Most of these works consider the evolution of an initially localized wavepacket in a chain of particles modeled by a continuous dynamical system and show that it spreads subdiffusively in moderately nonlinear systems, while for strong enough nonlinearities a substantial part of the wavepacket remains self-trapped. In particular, for single-site excitations, the spreading of the wavepacket leads to an increase of the second moment $m_2$ in time of the form $m_2 \sim t^{1/3}$, both in the diffusive and self-trapping regimes \cite{pikovsky_destruction_2008,Flach2009,Skokos2009,veksler_spreading_2009}.

Currently, a greatly debatable question concerns the long time behavior of wavepacket spreading in disordered nonlinear particle lattices. Recently, it was conjectured \cite{Johansson_kam_2010,A11} that chaotically spreading wavepackets will asymptotically approach Kolmogorov-Arnold-Moser (KAM) torus-like structures in phase-space, while in all numerical simulations chaotic spreading shows no sign of slowing down \cite{Laptyeva_crossover_2010,BLSKF11,SGF13}. Nevertheless, for particular disordered nonlinear models, some numerical indications of a possible slowing down of spreading have been reported \cite{PF11,MP13}. In Ref. \cite{Antonopoulosetal2014}, employing ideas from nonextensive statistical mechanics \cite{Tsallisbook2009}, we studied the same problem and found no such sign of quasi-periodic KAM behavior and all our evidence appears to strengthen the conjecture that waves spread subdiffusively and chaotically for arbitrarily long times in nonlinear disordered media. It would be interesting, therefore, if similar results could also be found in coupled nonlinear maps modeling nonlinear disordered media as they are computationally easier to study than the systems of ordinary differential equations of their Hamiltonian counterparts. Hamiltonian systems and symplectic maps, and in general discrete time systems, can be identified through discretization techniques, i.e. Poincar\'e sections. The drawback of this approach however is that the resulting map is in general, not explicitly given and thus one has to resort to the construction of a suitable discrete system that is able to exhibit similar dynamical and statistical properties as the disordered Hamiltonian counterpart.

In the present paper, we use tools of nonextensive statistical mechanics to investigate the connection between regimes of ``weak'' and ``strong'' chaos and diffusion in a system of $N$ coupled symplectic McMillan maps (CMM) \cite{McMillan1971,Glasseretal1989}, with the purpose of linking them to analogous results obtained recently for a Klein-Gordon (KG) Hamiltonian system in the presence of disorder and nonlinearity \cite{Antonopoulosetal2014}. In this framework, we are interested to explore the question whether coupled symplectic maps can be used to study subdiffusive spreading of wavepackets in disordered media, instead of using continuous flow dynamical models such as the KG Hamiltonian.

As is well-known, probability distribution functions (pdfs) of chaotic trajectories of dynamical systems have been studied by many authors, aiming to understand the transition from deterministic to stochastic dynamics \cite{Arnold1967,Sinai1972,Eckmann1985}. One of the most fundamental questions concerns the existence of an appropriate invariant probability density or ergodic measure, characterizing chaotic motion in phase-space. If it is possible to demonstrate the existence of such a measure, one can postulate that the system is at thermal equilibrium and can therefore be studied from the point of view of classical statistical mechanics.

Since the existence of an invariant measure is not known \textit{a priori}, one can still proceed in the context of the central limit theorem \cite{Rice1995} and consider the variables of a chaotic solution at discrete times $t_i,\;i=1,\ldots,M$ as realizations of $\mathcal{N}$ independent and identically distributed (iid) random variables $X^{(j)}(t_i),\;j=1,\ldots,\mathcal{N}$. If the motion is uniformly ergodic in some region of phase-space, one finds that the pdfs of the {\em sums} of these variables converge rapidly to a Gaussian distribution, whose mean and variance are those of the $X^{(j)}$'s. In such cases we call the dynamics ``strongly'' chaotic, since at least one Lyapunov exponent is positive \cite{Benettin1980a,Benettin1980b,S10} and the respective subset of the constant energy manifold is uniformly covered by chaotic orbits, for all but a (Lebesgue) measure zero set of initial conditions.

Now, if the system has not yet reached thermal equilibrium, it may happen that its orbits ``stick'' for long times on the boundaries of islands surrounding stable periodic orbits, where Lyapunov exponents \cite{Benettin1980a,Benettin1980b} become very small and may even vanish \cite{S10,Tsallisbook2009}. In such regimes, the motion may be called ``weakly'' chaotic, as trajectories get trapped within complicated sets of cantori and diffuse through multiply connected domains of phase-space in a highly nonuniform way \cite{Aizawa1984,Chirikov1984,Meiss1986,BouSkobook2012}. Many such examples have been discovered in physically realistic systems studied recently \cite{Skokos2008,Flach2009,Johansson2009,Skokos2009}.

In the present paper, we compare the diffusive dynamics of a system of coupled symplectic McMillan maps \cite{McMillan1971,Glasseretal1989}, with analogous results obtained recently in Ref. \cite{Antonopoulosetal2014} for a KG Hamiltonian 1-dimensional lattice in the presence of disorder and nonlinearity. In particular, we find in both models that weakly chaotic dynamics is associated with subdiffusive motion and is characterized by pdfs of sums of variables that \textit{do not} rapidly converge to a Gaussian, but are well approximated for long times by the so-called $q$-Gaussian distribution \cite{Tsallisbook2009}
\begin{equation}\label{q_gaussian}
\mathtt{P}(s)=a \exp_q({-\beta s^2})\equiv a\biggl[1-(1-q)\beta s^{2}\biggr]^{\frac{1}{1-q}},
\end{equation}
where the $q$ entropic index satisfies $1<q<3$, $\beta$ is an arbitrary parameter and $a$ is a normalization constant. Eventually, of course, it is  expected that chaotic orbits will wander within larger chaotic seas, where the dynamics is more strongly ergodic. When this occurs, the $q$ entropic index of Eq. (\ref{q_gaussian}) decreases towards $q=1$, which represents the limit at which the pdf becomes a Gaussian distribution (see Fig. \ref{distributions_N100_2_different_etas} for an illustrative example).

In Ref. \cite{Antonopoulosetal2014}, we studied wavepacket spreading in a Hamiltonian system represented by a disordered nonlinear KG chain of $N=1000$ particles \cite{BLSKF11,Laptyeva_crossover_2010} and found that even though there are phase-space regions of weak chaos, strongly chaotic dynamics eventually prevails characterized by $q$-Gaussian pdfs that approach a Gaussian pdf for long times. We have thus concluded that the motion will never approach a KAM regime of invariant tori as suggested by some authors \cite{Johansson_kam_2010,A11}.

In analogy with these findings, the statistics of diffusive motion in chains of $N$ CMM maps yield similar results to those for the KG system, although there are substantial differences in the dynamics of the two systems. In particular, in the KG system the dynamics is such that each particle has an elliptic point at the origin, while in our study for the CMM maps each map has a saddle point at the origin. Moreover, the KG system has a conserved quantity (see Eq. \eqref{RQKG}), while there is no such known quantity for the studied CMM maps of Eqs. \eqref{CMM_2012_09_07}. Another difference is that the KG system is characterized by diffusive processes of one epoch \cite{Flach2009} whereas in our study we find diffusion described by two different epochs. Finally, in the KG system large nonlinearities result in selftrapping, while smaller nonlinearities are connected to subdiffusion. In contrast, in the CMM maps large coupling strengths lead to diffusion while smaller values show no spreading. Despite these differences, we do find evidence of subdiffusion accompanied by weak and strong chaos, as in the case of the disordered KG chain, when the maps have a saddle point at their origin and the coupling strength is high enough. In that sense our results demonstrate that coupled symplectic maps can be used to model diffusive motion.

Our paper is organized as follows: In Sec. \ref{CLT_approach} we outline the details of our study of statistical distributions corresponding to weakly and strongly chaotic behavior, while in Sec. \ref{KG results} we summarize the results of Ref. \cite{Antonopoulosetal2014} to help the reader compare with our results on the CMM system presented in Sec. \ref{CMM}. Finally, Sec. \ref{conclusions} contains our conclusions and discussion.

\section{Computation of statistical distributions of weak and strong chaos}\label{CLT_approach}

In this section we focus our attention on the statistical properties of chaotic diffusion in a 2$N$-dimensional CMM system of $N$ linearly coupled 2-dimensional symplectic maps evolving in discrete time $n$. We wish to explore diffusion in regimes of weak chaos, where Lyapunov exponents are positive but very small. Such situations often arise when solutions move slowly through thin chaotic layers, wandering through a complicated network of higher order resonances, often sticking for very long times to the boundaries of islands constituting the so-called ``edge of chaos'' regime \cite{Tsallisbook2009}.

Many interesting questions can be asked in this context: How long do these weakly chaotic states last? What types of pdfs characterize their statistical behavior and how can one relate them to the diffusive properties of the motion? Is there a connection between complex statistics and diffusion as the dimensionality of the system increases? What is the effect of different parameters and what role does disorder play?

To answer these questions, we use the orbits of a multi-dimensional CMM model to construct pdfs of suitably rescaled sums of $M$ values of a generic observable $\eta_i=\eta(t_i),\;i=1,\ldots,M$ that depends linearly on the coordinates $x$ of the orbit. Viewing these as iid random variables in the limit of $M\rightarrow\infty$, we evaluate their sum
\begin{equation}\label{sums_CLT}
S_M^{(j)}=\sum_{i=1}^M\eta_i^{(j)}
\end{equation}
for $j=1,\ldots,N_{\mbox{ic}}$ different initial conditions and study the statistics of Eq. (\ref{sums_CLT}) centered about their mean value $\langle S_M^{(j)}\rangle=\frac{1}{N_{\mbox{ic}}}\sum_{j=1}^{N_{\mbox{ic}}}\sum_{i=1}^{M}\eta_i^{(j)}$ and rescaled by their standard deviation $\sigma_M$
\begin{equation}
s_M^{(j)}\equiv\frac{1}{\sigma_M}\Bigl(S_M^{(j)}-\langle S_M^{(j)}\rangle \Bigr)=\frac{1}{\sigma_M}\Biggl(\sum_{i=1}^M\eta_i^{(j)}-\frac{1}{N_{\mbox{ic}}}\sum_{j=1}^{N_{\mbox{ic}}}\sum_{i=1}^{M}\eta_i^{(j)}\Biggl),
\end{equation}
where
\begin{equation}
\sigma_M^2=\frac{1}{N_{\mbox{ic}}}\sum_{j=1}^{N_{\mbox{ic}}}\Bigl(S_M^{(j)}-\langle S_M^{(j)}\rangle \Bigr)^2=\langle S_M^{(j)2}\rangle -\langle S_M^{(j)}\rangle^2.
\end{equation}
Plotting the normalized histogram of the probabilities $\mathtt{P}(s_M^{(j)})$ as a function of $s_M^{(j)}$, we compare our pdfs with a $q$-Gaussian of the form
\begin{equation}
\mathtt{P}(s_M^{(j)})=a\exp_q({-\beta s_M^{(j)2}})\equiv a\biggl[1-(1-q)\beta s_M^{(j)2}\biggr]^{\frac{1}{1-q}},
\end{equation}
cf. (\ref{q_gaussian}), where $q$ is the so-called entropic index. Note that this is a generalization of the well-known Gaussian pdf, since in the limit $q\rightarrow 1$ we have $\exp_q(-\beta x^2)\rightarrow\exp(-\beta x^2)$. Moreover, it can be shown that the $q$-Gaussian distribution (\ref{q_gaussian}) is normalized when
\begin{equation}\label{beta-$q$-Gaussian}
\beta=a\sqrt{\pi}\frac{\Gamma\Bigl(\frac{3-q}{2(q-1)}\Bigr)}{(q-1)^{\frac{1}{2}}\Gamma\Bigl(\frac{1}{q-1}\Bigr)},
\end{equation}
where $\Gamma$ is the Euler $\Gamma$ function. Clearly, Eq. (\ref{beta-$q$-Gaussian}) shows that the allowed values of $q$ are $1<q<3$ for this normalization.

The index $q$ appearing in Eq. (\ref{q_gaussian}) is connected with the Tsallis entropy \cite{Tsallisbook2009}
\begin{equation}\label{Tsallis entropy}
S_q=k\frac{1-\sum_{i=1}^W \mathcal{P}_i^q}{q-1}\mbox{ with }\sum_{i=1}^W \mathcal{P}_i=1,
\end{equation}
where $i=1,\ldots,W$ counts the microstates of the system, each occurring with a probability $\mathcal{P}_i$ and $k$ is the well-known Boltzmann constant. Just as the Gaussian distribution represents an extremal of the Boltzmann-Gibbs entropy $S_{\mbox{BG}}\equiv S_1=k\sum_{i=1}^W \mathcal{P}_i\ln \mathcal{P}_i$, so is the $q$-Gaussian (\ref{q_gaussian}) derived by optimizing the Tsallis entropy of Eq. (\ref{Tsallis entropy}) under appropriate constraints.

Systems characterized by the Tsallis entropy are said to lie at the ``edge of chaos'' and are significantly different from Boltzmann-Gibbs systems, in the sense that their entropy is nonadditive and generally nonextensive \cite{Tsallisbook2009}. In fact, a $q$-central limit theorem has been proved \cite{UmarovTsallis2008} for $q$-Gaussian distributions (\ref{q_gaussian}) that is of the same form as the classical central limit theorem.

As we demonstrate in the following sections, in regions of weak chaos these distributions are well-fitted by a $q$-Gaussian pdf for fairly long time intervals, whose $q$ value is distinctly greater than 1. It may happen, of course, for longer times that the orbits begin to diffuse through domains of strong chaos, in which case $q$ tends to 1 and the well-known form of a Gaussian pdf is eventually recovered.

To study diffusive processes of chaotic dynamics in multi-dimensional CMMs, we adopt the classical definition of the mean square displacement (MSD) $\langle r^2\rangle$ given by the normalized sum of the squares of the distances of the $N_{\mbox{ic}}$ orbits from their starting point. If the motion is a random walk, we expect that
\begin{equation}
\langle r^2\rangle = D n^{\gamma},
\end{equation}
where $n$ denotes the iterations of the dynamics, $\gamma<1$ characterizes the process as subdiffusive, $\gamma=1$ corresponds to normal diffusion and $\gamma>1$ to super-diffusive motion. $D$ is the diffusion coefficient and $\gamma=2$ corresponds to ballistic motion. Finally, we compute the corresponding Lyapunov exponents of the dynamics following the well-known methodology presented in Refs. \cite{Benettin1980a,Benettin1980b,S10}.

\section{Chaotic diffusion in a KG particle chain}
\label{KG results}

In the case of a high-dimensional KG disordered particle chain, one may concentrate on a low energy (subdiffusive) and a higher energy (self-trapping) case and verify that subdiffusive spreading always occurs following specific power-laws. Integrating the equations of motion for long times and computing probability distributions of sums of the positions of particles, we have found in Ref. \cite{Antonopoulosetal2014} convincing evidence that the dynamics does not relax onto a quasi-periodic KAM torus-like structure, but continues to spread chaotically along the KG chain of particles for arbitrarily long times.

To show this, we considered a quartic KG lattice described by the $N$ degree of freedom Hamiltonian
\begin{equation}
H_{KG}= \sum_{l=1}^{N} \frac{p_{l}^2}{2} + \frac{\tilde{\epsilon}_{l}}{2}
x_{l}^2 + \frac{1}{4} x_{l}^{4}+\frac{1}{2W}(x_{l+1}-x_l)^2=E,
\label{RQKG}
\end{equation}
where $x_l$ and $p_l$ are the generalized positions and momenta on site $l$, and $\tilde{\epsilon}_{l}$ are chosen uniformly randomly from the interval $\left[\frac{1}{2},\frac{3}{2}\right]$ to account for the disorder present at each site $l$ of the chain. This Hamiltonian conserves the value of the total energy $E\geq 0$ of the system, which, for fixed disorder strength $W$, serves as a control parameter of the nonlinearity. In our study, we followed the evolution of single site excitations solving the equations of motion
\begin{equation}		
\ddot{x}_{l} = - \tilde{\epsilon}_{l}x_{l} -x_{l}^{3} + \frac{1}{W}
(x_{l+1}+x_{l-1}-2x_l),\;l=1,\ldots,N,
\label{KG-EOM}
\end{equation}
with fixed boundary conditions (i.e. $x_0=x_{N+1}=0$) and monitored the corresponding normalized energy density distributions.

\begin{figure}[!ht]
\centering{
\includegraphics[scale=0.85]{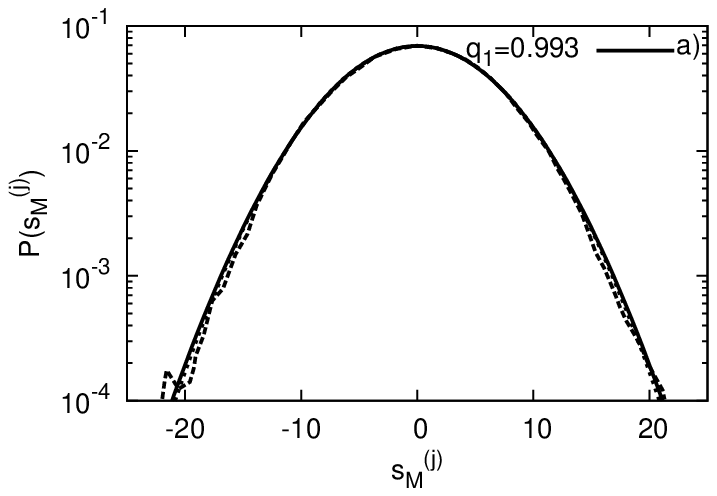}
\includegraphics[scale=0.85]{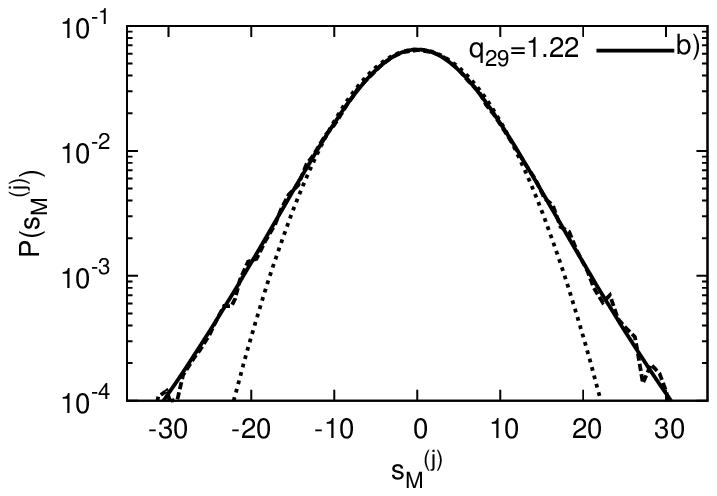}
\caption{\textbf{Characteristic pdfs for the disordered $\boldsymbol{N=1000}$ KG system for single site excitations.} Panel a): Plot of the numerically computed pdf (dashed curve) for the observable $\eta_1=x_{500}$ in the time interval $[0,10^8]$ with $q_1=0.993 \pm 0.009$, taken from fitting with a $q$-Gaussian distribution (\ref{q_gaussian}) in solid thick. Panel b): Similar plot of the numerically computed pdf (dashed) for the observable $\eta_{29}=\sum_{l=486}^{514}x_{l}$ and a time interval $[0,10^8]$ with $q_{29}=1.22 \pm 0.01$, taken from fitting with a $q$-Gaussian distribution (\ref{q_gaussian}) in solid thick. In both panels, $N=1000$, $M=10^7$ and $E=0.4$ that corresponds to the subdiffusive case. Note that the vertical axes are in logarithmic scale, while the dotted curve is the Gaussian pdf (i.e. $q=1$).\newline Reproduced from Chaos 24, 024405 (2014).\newline\copyright~2014 AIP}\label{distributions_N1000_E0.4_different_etas}}
\end{figure}

In Fig. \ref{distributions_N1000_E0.4_different_etas} (taken as is from Ref. \cite{Antonopoulosetal2014}) we exhibit two representative examples of numerical pdfs with different $q$ entropic indices for a low energy subdiffusive case with $E=0.4$ and $N=1000$ (the initial value of $x_{500}(0)$ is adjusted so that $E=0.4$). In panel a) we plot the numerical distribution (dashed curve) for the observable $\eta_1=x_{500}$ computed in the time interval $[0,10^8]$ for $M=10^7$ and note that it is well fitted by a $q$-Gaussian distribution (solid thick curve) with $q_1=0.993 \pm 0.009$. This is a case where the numerical distribution is indistinguishable from a Gaussian ($q=1$) plotted as a dotted curve. On the other hand, panel b) which is the same plot as a) for the observable $\eta_{29}=\sum_{l=486}^{514}x_{l}$, reveals a clear $q$-Gaussian distribution (\ref{q_gaussian}) with $q_{29}=1.22 \pm 0.01$, over nearly four decades on the vertical axis.

In this way, we studied in Ref. \cite{Antonopoulosetal2014} the dynamics and statistics of diffusive motion in the 1-dimensional nonlinear disordered KG particle chain and obtained results that can be directly compared to what we describe in the present paper using CMM maps. In particular, for a disordered KG chain of $N=1000$ particles, we focused on a low energy (subdiffusive) and a higher energy (self-trapping) case and verified that subdiffusive spreading always occurs following specific power-laws with exponents smaller than 1 as pointed out in the literature \cite{Flach2009}. Moreover, integrating the equations of motion for times as long as $10^9$ and computing the corresponding pdfs, we found evidence that the dynamics does {\em not} relax onto a quasi-periodic KAM torus, as it has been conjectured \cite{Johansson_kam_2010,A11}, but continues to spread chaotically along the chain for arbitrarily long times.

\section{Chaotic diffusion in chains of coupled symplectic maps}
\label{CMM}

Typically, Hamiltonian systems and in general discrete time dynamical systems can be analyzed with the help of discretization techniques, such as Poincar\'e sections. The disadvantage is that the resulting map is, in general, not explicitly known and thus one has to resort to the construction of a suitable discrete system that exhibits similar statistical properties with its disordered Hamiltonian counterpart. Moreover, since the numerical integration of Hamilton's equations for long particle chains is often too costly, one might wonder whether it is possible to use a system of symplectic maps to model the dynamics and statistics of multi-degree of freedom Hamiltonian systems, especially in the physically interesting case where nonlinearity and disorder coexist \cite{Flach2009,Skokos2009,BLSKF11,Laptyeva_crossover_2010}. This would mean that each particle in such a chain corresponds to a 2-degree of freedom oscillator and is modeled by a 2-dimensional symplectic map.

We decided therefore to describe each particle by the 2-dimensional McMillan map \cite{Glasseretal1989} given by
\begin{equation}
\label{McMillanMap}
x_{n+1}=x_n+p_{n+1},\qquad p_{n+1}=p_n+\frac{2Kx_n}{x_n^2+1}-2x_n.
\end{equation}
This map is integrable, possessing the constant of the motion $I_n=x_n^2y_n^2+x_n^2+y_n^2-2Kx_ny_n$, with $I_n=I_{n+1}$ for all $n=0,1,2,\ldots$ and $y_n=x_n-p_n$. If $|K|<1$, the origin in the $(x_n,p_n)$ plane is elliptic, while it becomes a saddle point via a pitchfork bifurcation for $K>1$ and via a period-doubling bifurcation for $K<-1$. In Ref. \cite{RuizBountisTsallis2012}, the authors studied a non-integrable conservative perturbation of map (\ref{McMillanMap}) and chaotic orbits were investigated from a statistical point of view, similar to the approach used here.

In our study, we shall consider a system of CMMs expressed in the form
\begin{equation}\label{CMM_2012_09_07}
	x_{n+1}^{(j)}=x_n^{(j)}+p_{n+1}^{(j)},\qquad
	p_{n+1}^{(j)}=p_n^{(j)}-\frac{\partial V}{\partial x_n^{(j)}},\;j=1,\ldots,N,
\end{equation}
where $x_n^{(j)}$, $p_n^{(j)}$ denote the position and momentum of the $j$-th map respectively, at the $n$-th step in discrete time (i.e. iteration). Thus, CMMs represent a discrete non-linear system of $N$ 2-dimensional symplectic McMillan maps with a quadratic nearest neighbor interactions potential of the form
\begin{equation}\label{potential_subsection_Results_2012_09_07}
	 V=\sum_{j=1}^N\Bigl(-K^{(j)}\ln(x_n^{(j)2}+1)+x_n^{(j)2}+\frac{\epsilon}{2}(x_n^{(j+1)}-x_n^{(j)})^2\Bigr),
\end{equation}
under fixed boundary conditions $x^{(0)}=x^{(N+1)}=0$, while $\epsilon\neq0$ is the coupling parameter that renders the system non-integrable.

\begin{figure}[!ht]
\centering{
\includegraphics[scale=0.55]{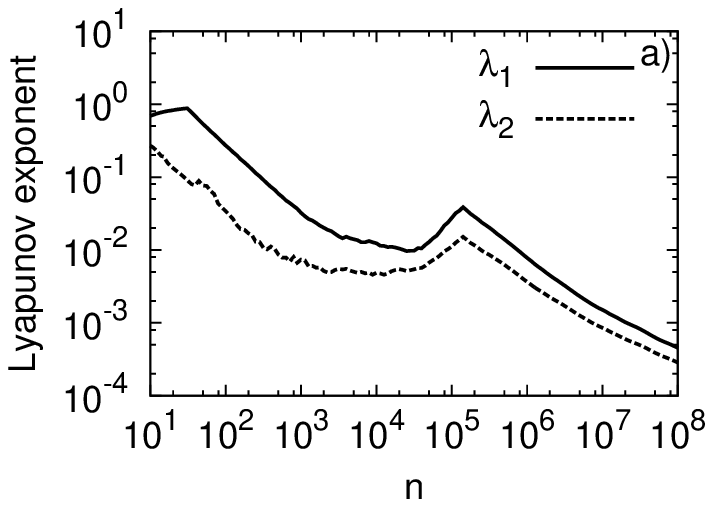}
\includegraphics[scale=0.55]{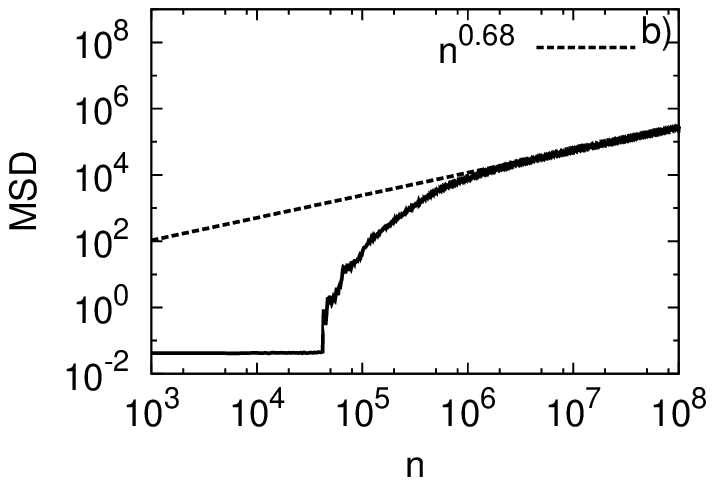}
\includegraphics[scale=0.55]{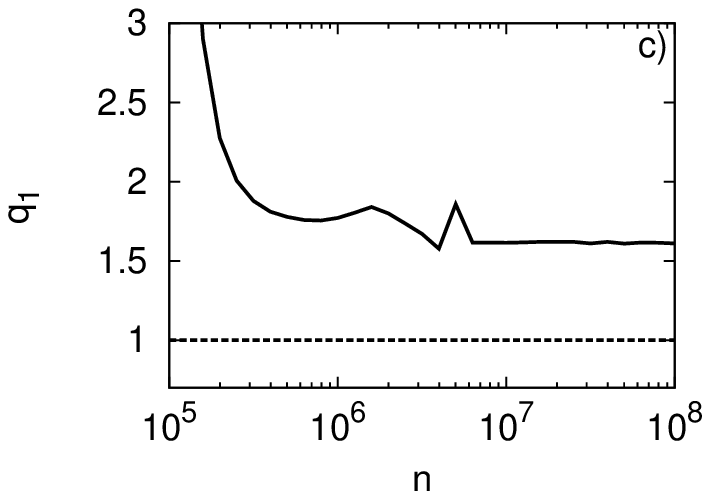}\\

\includegraphics[scale=0.55]{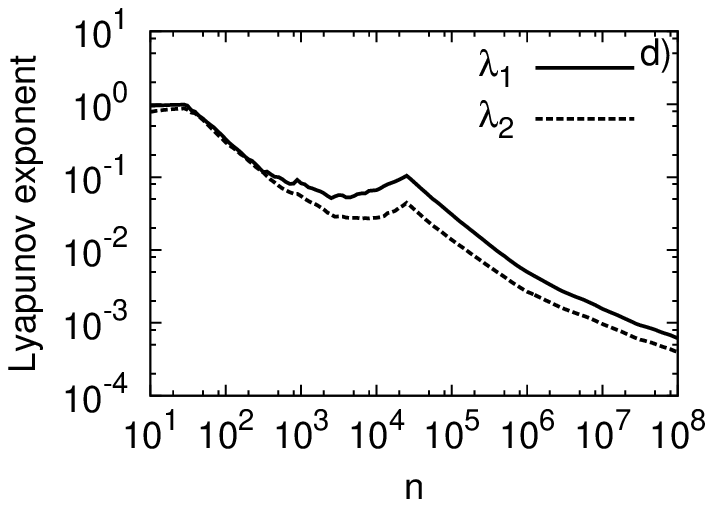}
\includegraphics[scale=0.55]{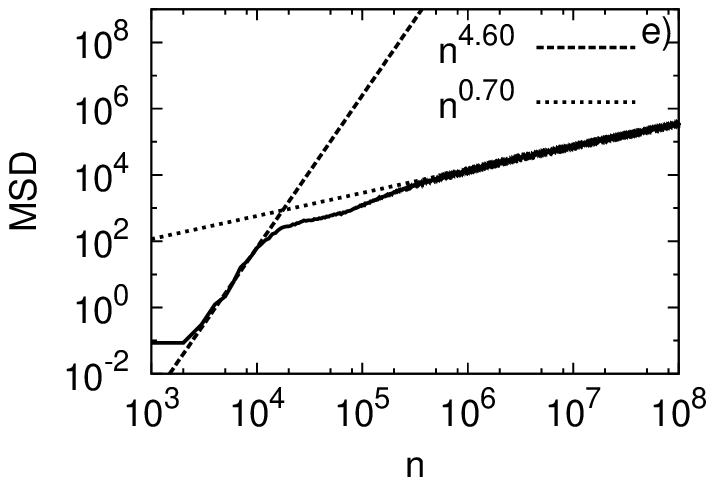}
\includegraphics[scale=0.55]{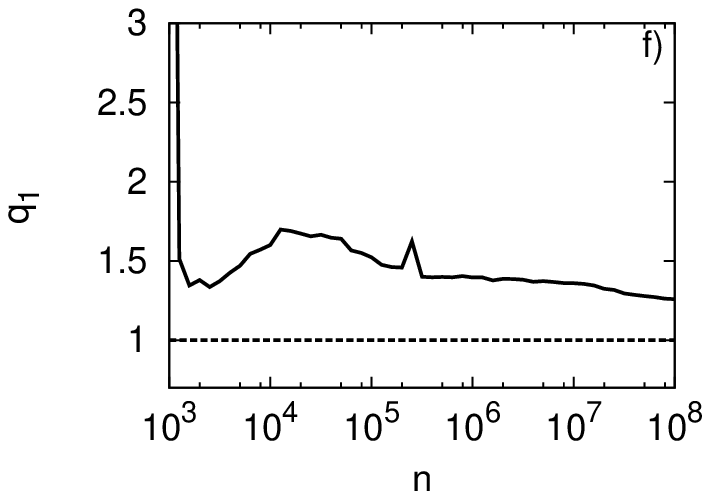}\\

\includegraphics[scale=0.55]{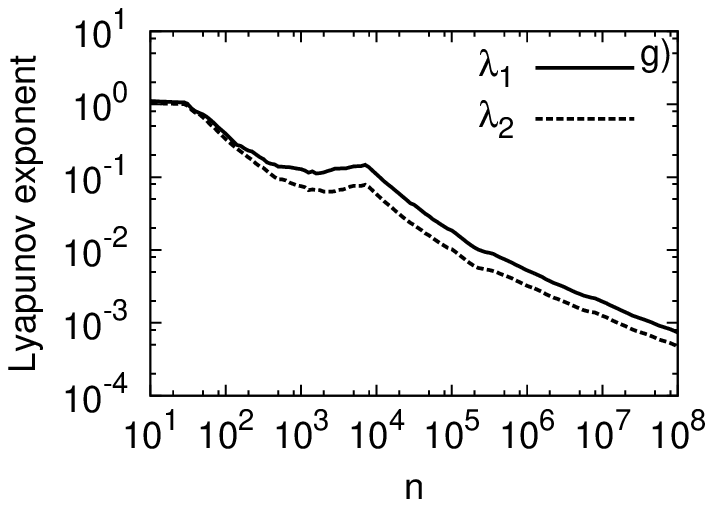}
\includegraphics[scale=0.55]{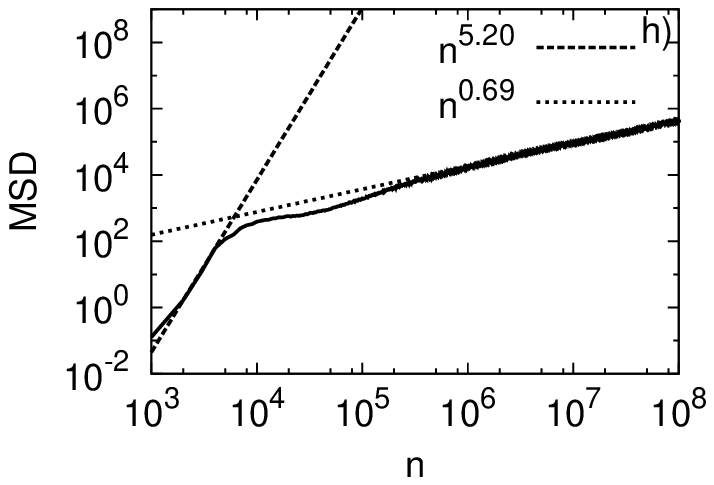}
\includegraphics[scale=0.55]{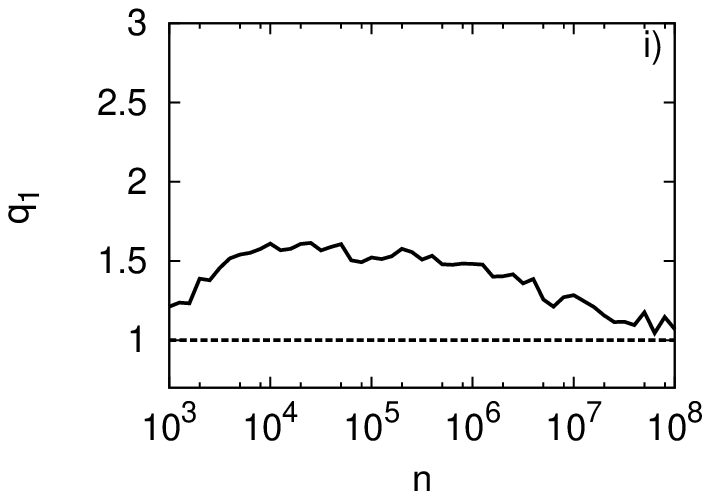}\\

\includegraphics[scale=0.55]{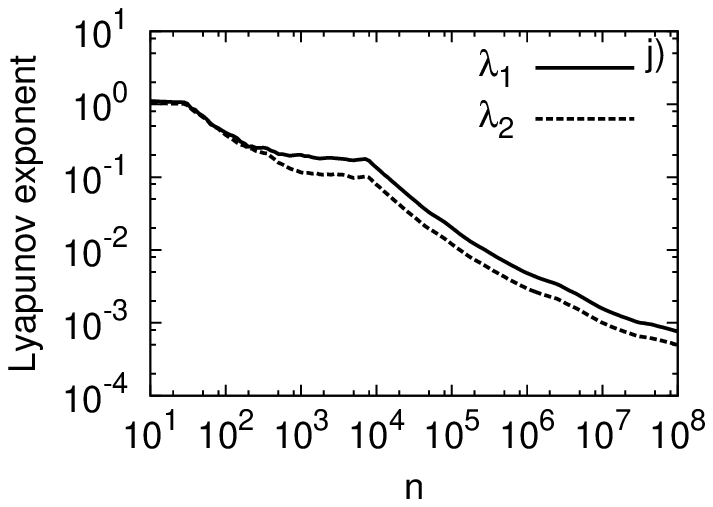}
\includegraphics[scale=0.55]{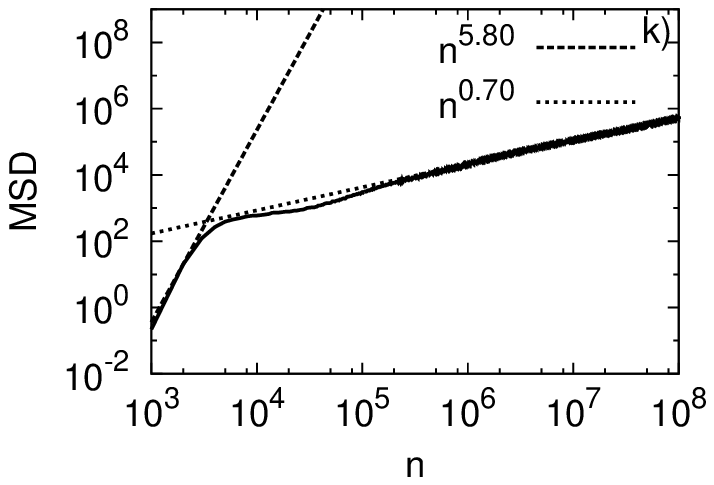}
\includegraphics[scale=0.55]{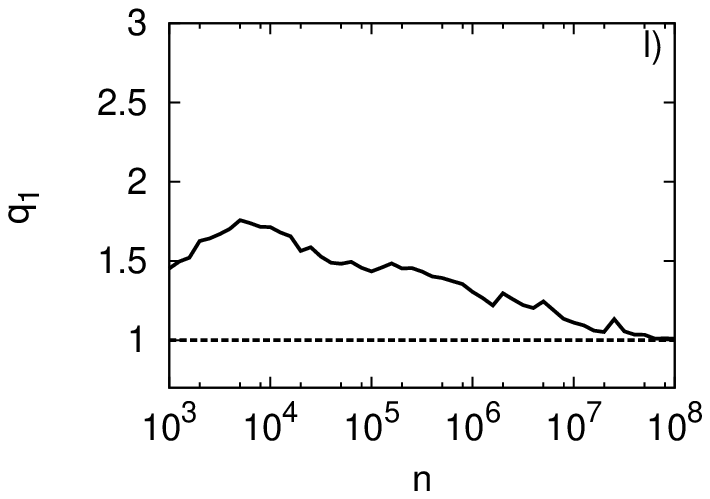}\\

\includegraphics[scale=0.55]{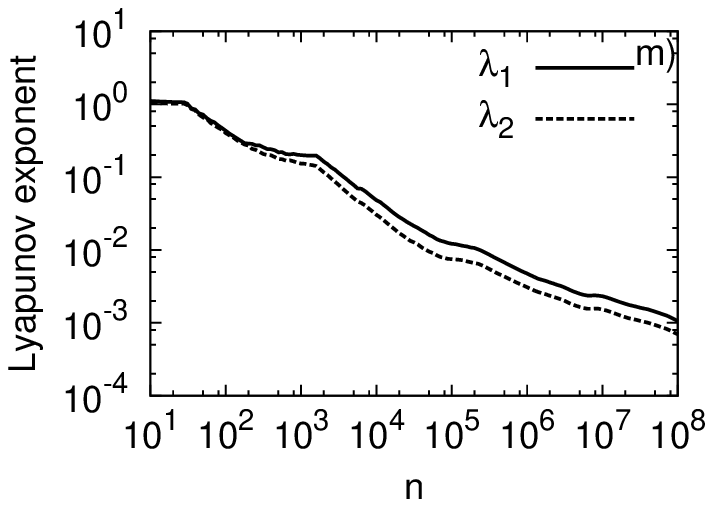}
\includegraphics[scale=0.55]{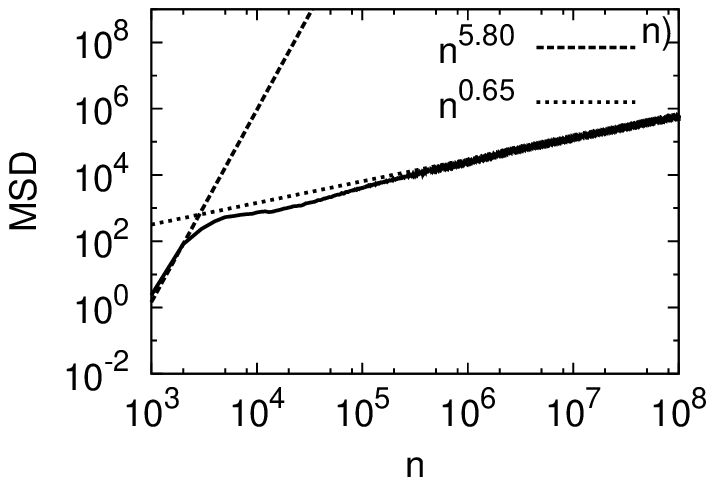}
\includegraphics[scale=0.55]{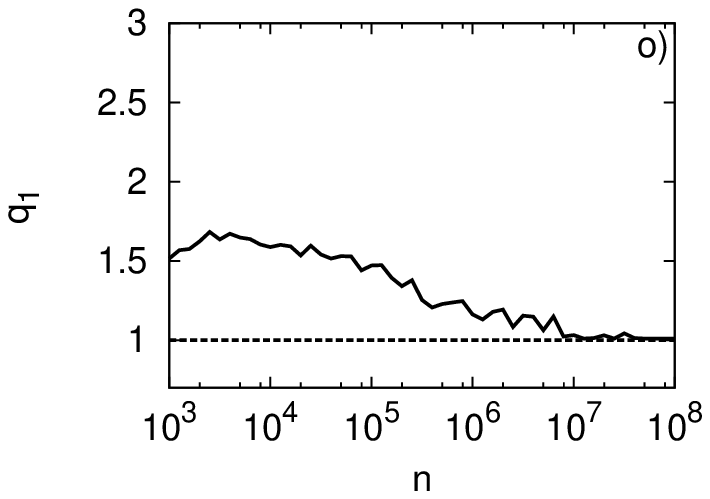}
\caption{\textbf{Results for the case of 5 CMMs with $\boldsymbol{\epsilon=0.1}$.} For each row, from left to right, we plot the corresponding two largest Lyapunov exponents $\lambda_1$, $\lambda_2$ (left panels), the MSD (central panels) and $q_1$ (right panels) as a function of $n$. The first row corresponds to $K^{(3)}=1.6$ and the rest of $K$'s equal to $0.6$, the second row to the case $K^{(2)}=K^{(3)}=1.6$ with all other $K$'s equal to $0.6$, the third panel to the case $K^{(1)}=K^{(2)}=K^{(3)}=1.6$ with the remaining $K$'s equal to $0.6$, the forth row to the case $K^{(1)}=K^{(2)}=K^{(3)}=K^{(4)}=1.6$ and $K^{(5)}=0.6$, and finally the fifth row to the case where $K^{(j)}=1.6,\;j=1,\ldots,5$. In all cases, $q_1$ corresponds to $n_1=x^{(3)}$ and $M=10^7$.}\label{fig_subsection_Results_2013_01_06}}
\end{figure}

In the uncoupled case (i.e. $\epsilon=0$), if $K^{(j)}>1,\;j=1,\ldots,N$ each map has a saddle point at its origin, while for $0<K^{(j)}<1$ the origin is elliptic. Choosing $K^{(j)}$ larger than 1, if $\epsilon>0$ is small enough, we make sure that the coupling does not alter the stability properties of the origin, while the dimensionality of its stable and unstable manifolds can still be adjusted by only varying the values of the $K^{(j)}$ parameters. For these reasons, we have chosen to use throughout the paper $K^{(j)}$ values in $[1.2,1.6]$ to emulate disorder as we have additionally checked by numerical simulations that any other choice of $K^{(j)}>1$ produces similar results. Finally, a preliminary study for different initial conditions (i.e. initial excitation strengths) revealed that as long as they are chosen sufficiently close to the origins of all maps, the obtained results are similar. This is supported by the fact that for $K^{(j)}>1$ all origins are saddle points with stable and unstable manifolds and thus a typical initial excitation strength of the $j$-th map $x_0^{(j)}=10^{-6}$ suffices to ensure that the above properties hold.

Now, if one starts with all maps having stable origins (all $K^{(j)}<1$) and excites only the central map, just as we did in Ref. \cite{Antonopoulosetal2014} for single-site excitations of the KG system, no chaotic diffusion is observed. We, therefore, turn to the more challenging scenario where the central map has a saddle point at its origin and use an ensemble of $N_{\mbox{ic}}$ initial conditions near that point, assuming that all other maps have elliptic points at their origin. Finally, we increase the number of maps with saddle point origins and study the dynamics and diffusion of orbits in phase-space by computing the corresponding pdfs of sums of their position coordinates $x_n$ as described in Sec. \ref{CLT_approach} to understand if there is any connection between chaotic diffusion in the two models.

We first considered $N=5$ coupled McMillan maps and studied the dynamics around the origin when the central map (i.e. $j=3$) is unstable. In the calculations of the pdfs, we used as an observable variable $\eta_1=x^{(3)}$ of the central map and averaged over $N_{\mbox{ic}}=10^4$ initial conditions randomly chosen within $[-10^{-10},10^{-10}]$ about a reference orbit started close to the origin. For $\epsilon$ as small as 0.001 we observed no diffusion since the corresponding $\gamma$ exponent was practically 0 and the pdfs of the sum variable $S_M=\sum_{i=1}^{M}x_i^{(3)}$ of the central particle were well approximated by $q$-Gaussians, whose index fluctuates around $q_1=2$ even after as many as $10^8$ iterations!

On the other hand, increasing the coupling strength to $\epsilon=0.1$ allows for the presence of stronger chaos and yields signs of diffusive behavior. In Fig. \ref{fig_subsection_Results_2013_01_06}, we show for the $N=5$ case (with $M=10^7$), in each row, from left to right, the two largest Lyapunov exponents $\lambda_1$, $\lambda_2$ (left panels), the MSD as a function of the number of iterations $n$ (central panels) and the $q_1$ entropic index of the central map (i.e. $n_1=x^{(3)}$) as a function of $n$ (right panels). Note that, as the number of unstable maps increases, $q_1$ approaches 1 with increasing $n$ and a diffusion process is observed whose exponents $\gamma$ for large enough $n$ are clearly smaller than 1. In particular, when more than two maps are unstable, there is first a transient state of rapid diffusion with MSD $\propto n^\gamma$ and $\gamma>2$, faster than ballistic, over a relatively small number of iterations. Then, as $n$ increases, a second epoch occurs with $\gamma<1$, corresponding to subdiffusive dynamical processes.

\begin{figure}[!ht]
\centering{
\includegraphics[scale=0.9]{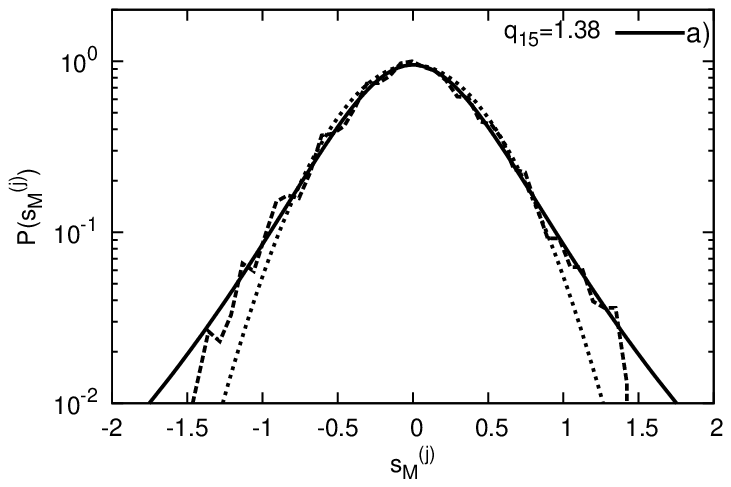}
\includegraphics[scale=0.9]{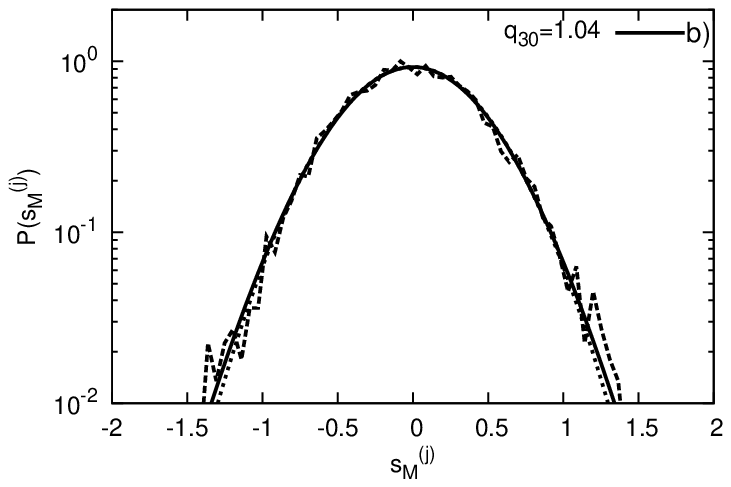}
\caption{\textbf{Characteristic pdfs for 30 disordered CMMs for single site excitations.} Panel a): Plot of the computed pdf (dashed) for the observable $\eta_{15}=\sum_{j=8}^{22}x^{(j)}$ (that represents the 15 central maps) for the time interval $[0,10^6]$ with $q_{15}=1.38 \pm 0.05$, taken from fitting with a $q$-Gaussian distribution (\ref{q_gaussian}) in solid thick. Panel b): Similar plot of the computed pdf (dashed curve) for the observable $\eta_{30}=\sum_{j=1}^{30}x^{(j)}$ (all maps) computed for the time interval $[0,10^8]$ with $q_{30}=1.04 \pm 0.05$, taken from fitting with a $q$-Gaussian distribution (\ref{q_gaussian}) in solid thick. In both panels, $\epsilon=0.28$, $N=30$, $M=5\cdot10^6$ and $K$ values were chosen randomly in $[1.2,1.6]$ to implement disorder. Both cases were taken from the second panel of Fig. \ref{fig_subsection_Results_2014_04_07}. Note that the vertical axes are in logarithmic scale, while the dotted curve is the Gaussian pdf with $q=1$ to guide the eye.}\label{distributions_N100_2_different_etas}}
\end{figure}

The next step is to set $N=30$ and increase the coupling further setting $\epsilon=0.28$ with single site excitations perturbing only the central map, and all $K^{(j)}$ values chosen randomly in the interval $[1.2,1.6]$ to emulate disorder as in the KG system. In this case, we obtain much more interesting statistics. As can be seen in Fig. \ref{distributions_N100_2_different_etas}, strong chaos is displayed when all maps are taken into account (panel b)) with $q_{30}=1.04\pm0.05$ (that corresponds to $\eta_{30}=\sum_{j=1}^{30}x^{(j)}$), while the 15 central maps of the chain exhibit weak chaos (panel a)) with $q_{15}=1.38\pm0.05$ (where $\eta_{15}=\sum_{j=8}^{22}x^{(j)}$). These findings are important as they demonstrate a similar situation to the physical problem of disordered KG coupled oscillators studied in Ref. \cite{Antonopoulosetal2014}, where the central particles perform weakly chaotic motion, while the overall system displays strongly chaotic dynamics depicted by their corresponding entropic indices.

\begin{figure}[!ht]
\centering{
\includegraphics[scale=0.85]{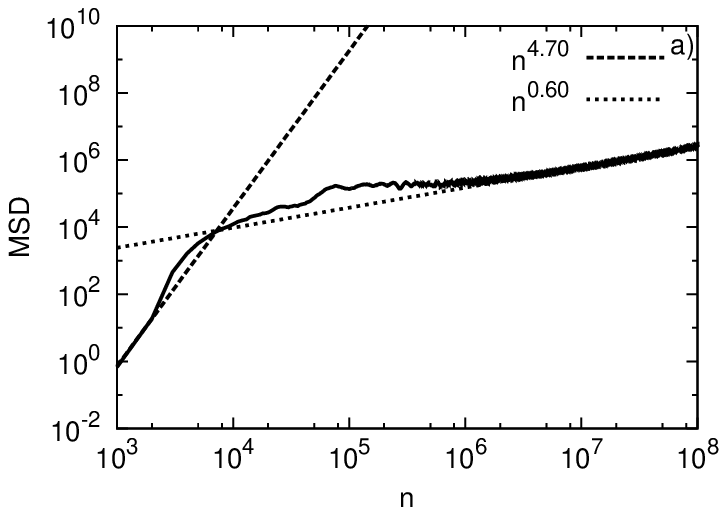}
\includegraphics[scale=0.85]{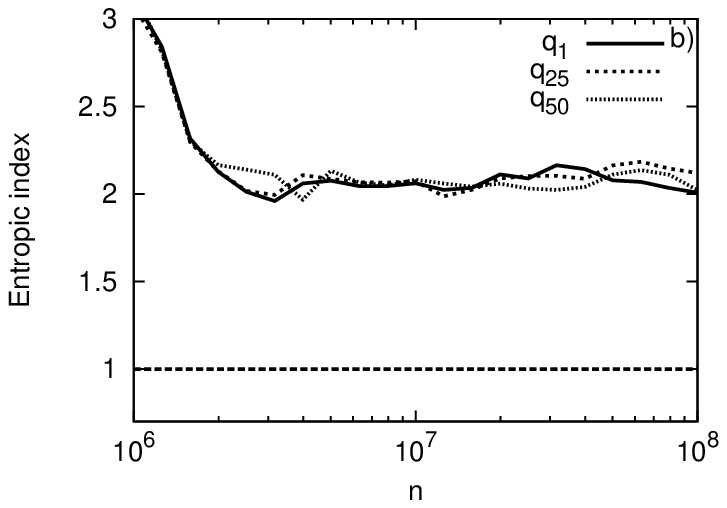}\\\includegraphics[scale=0.85]{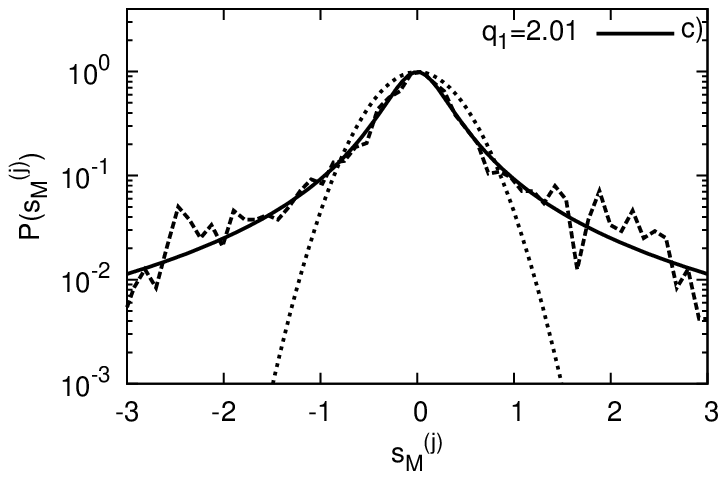}
\includegraphics[scale=0.85]{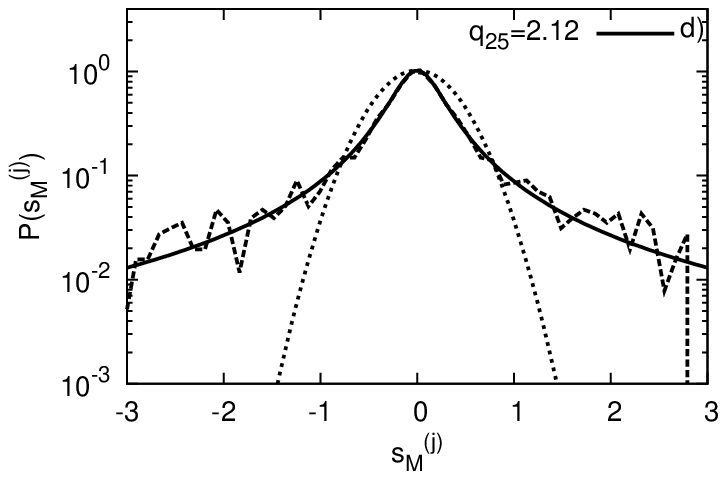}
\includegraphics[scale=0.85]{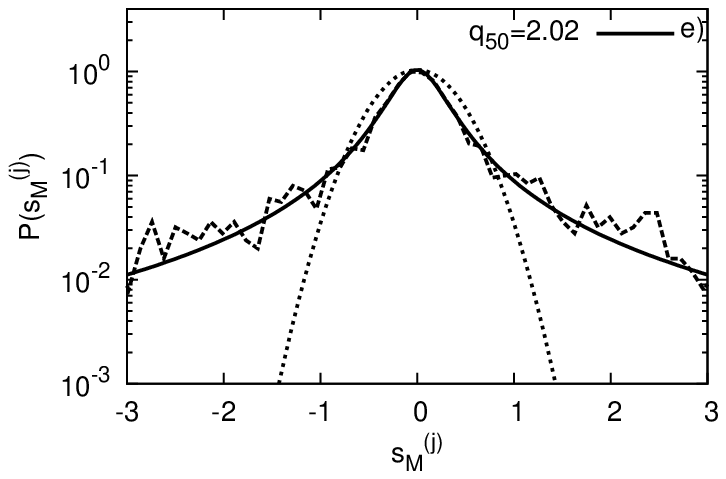}
\caption{\textbf{Results for the case of 50 disordered CMMs and single site excitations with $\boldsymbol{\epsilon=0.1}$.} Panel a): Plot of the MSD as a function of $n$ for $N_{\mbox{ic}}=2\cdot10^3$, all maps locally unstable by using $K$ values randomly chosen in $[1.2,1.6]$ with only the central map initially excited. Panel b) shows the evolution of the $q$ entropic index as a function of $n$ for the central map ($q_1$ with $\eta_1=x^{(25)}$), the 25 central maps ($q_{25}$ with $\eta_{25}=\sum_{j=13}^{37}x^{(j)}$) and all 50 maps ($q_{50}$ for $\eta_{50}=\sum_{j=1}^{50}x^{(j)}$) considered. The computed pdf (dashed curve), the fitted $q$-Gaussian (solid thick curve) of Eq. (\ref{q_gaussian}) and the Gaussian pdf (dotted curve) are shown at $n=10^8$ iterations: Panel c) for the central map, d) for the 25 central maps and e) for all 50 maps. In all panels for the entropic index $q$, we have used $M=5\cdot10^6$.}
\label{fig_subsection_Results_2013_06_30_N=50_allKin1.2_1.6_0.1}}
\end{figure}

Based on these first results, our next step was to investigate CMMs of increasing $\epsilon$ to relate complex statistics with diffusive dynamics. We did that for 30, 50 and 100 disordered CMMs using single site excitations. In all cases studied, we witnessed the same scenario: In order to observe (sub)diffusion in the disordered CMM, the coupling strength had to be relatively large ($\epsilon>0.05$) with all maps being locally unstable at the origin with all $K^{(j)}$ randomly distributed in $[1.2,1.6]$. We thus concluded that, for small enough coupling, orbits do not diffuse in phase-space but rather remain wandering around the unstable origins of each CMM. This behavior was clearly depicted by the entropic indices as the central particle performed a weakly chaotic motion with $q_1$ values substantially higher than unity whereas the rest of the chain behaved in an essentially strongly chaotic manner approaching $q\approx1$ in the long run.

We then turned to bigger values of $\epsilon$, focusing on 30 and 50 CMMs as the case of 100 CMM performed similarly. In particular, for 50 CMMs and $\epsilon=0.1$, exciting only the central map and taking all maps to have a saddle point at their origin, we found that after an initial number of about $10^4$ iterations (during which the orbits diffuse very fast with MSD $\propto n^{4.7}$), the chain settles down to a clearly subdiffusive regime with MSD $\propto n^{0.6}$ as evidenced in the first panel of Fig. \ref{fig_subsection_Results_2013_06_30_N=50_allKin1.2_1.6_0.1}. This behavior, however, is associated with weak chaos, since the entropic indices of panels b) to e) of the same figure, for the central map ($q_1$ for $\eta_{1}=x^{(25)}$), the 25 central maps ($q_{25}$ for $\eta_{25}=\sum_{j=13}^{37}x^{(j)}$) and the whole chain ($q_{50}$ for $\eta_{50}=\sum_{j=1}^{50}x^{(j)}$) are as high as 2 for large $n$!

\begin{figure}[!ht]
\centering{
\includegraphics[scale=0.85]{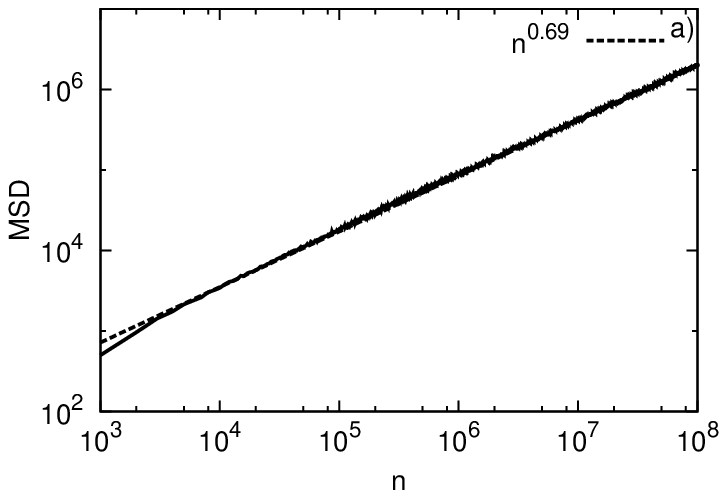}
\includegraphics[scale=0.85]{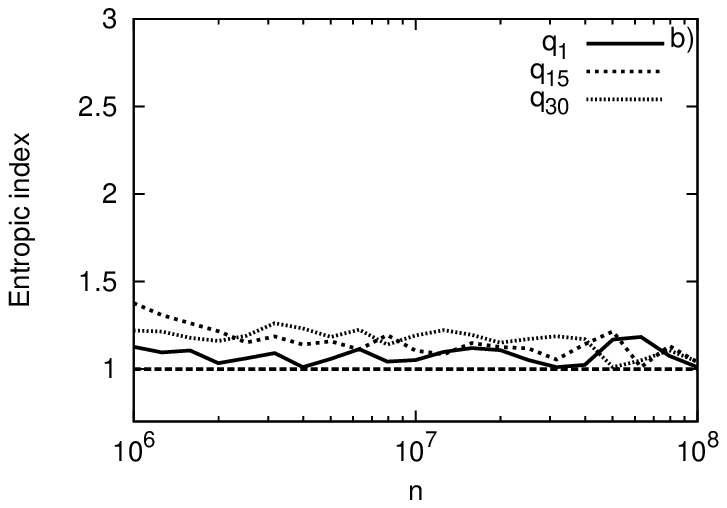}\\
\includegraphics[scale=0.85]{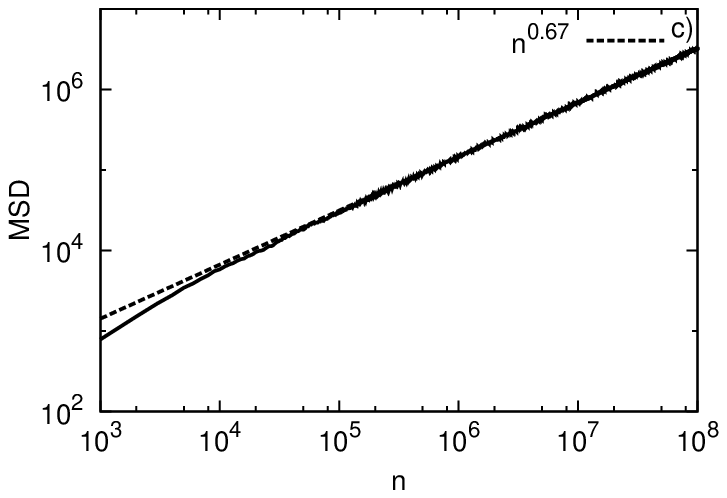}
\includegraphics[scale=0.85]{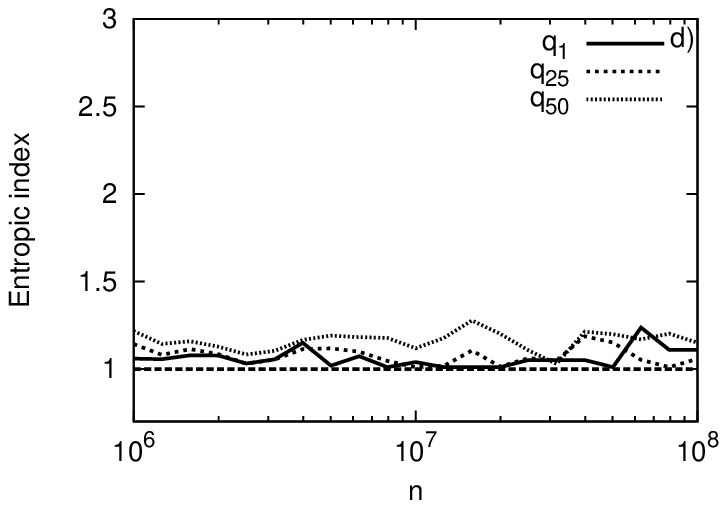}
\caption{\textbf{Subdiffusive chaotic behavior of large CMM disordered systems ($\boldsymbol{N=30}$ and 50).} Panel a): Plot of the MSD as a function of $n$ for $N=30$ CMMs, $\epsilon=0.28$, $N_{\mbox{ic}}=4\cdot10^3$, all maps locally unstable and $K$ values randomly chosen in $[1.2,1.6]$ with only the central one initially excited. Panel b) shows the evolution of the $q$ entropic index as a function of $n$ for the central map ($q_1$ for $\eta_{1}=x^{(15)}$), the 15 central ones ($q_{15}$ for $\eta_{15}=\sum_{j=8}^{22}x^{(j)}$) and all maps ($q_{30}$ for $\eta_{30}=\sum_{j=1}^{30}x^{(j)}$) when considering $N=30$ CMMs. Panels c) and d) are similar for $N=50$, $\epsilon=0.29$ and $N_{\mbox{ic}}=2\cdot10^3$. Note that all horizontal and vertical axes of the MSD plots are logarithmic. In all panels for the entropic index $q$, we have used $M=5\cdot10^6$.}
\label{fig_subsection_Results_2014_04_07}}
\end{figure}

Using then coupling strengths as high as 0.28 for 30 and 0.29 for 50 CMMs, presented in Fig. \ref{fig_subsection_Results_2014_04_07}, we did observe fully chaotic behavior for all different parts of the chain accompanied by subdiffusive motion. This is illustrated in the first two panels of this figure for 30 CMMs and the last two for 50 CMMs. Clearly, in the long run, the subdiffusive motion is characterized by an exponent $\gamma\approx0.68$ and by all entropic indices fluctuating in the interval $[1,1.2]$, signaling that the dynamics of the chain is close to fully developed chaos. These findings provide evidence that the CMM system in the long run, for sufficiently high dimensionality and coupling strength,sup does show evidence of strong chaos and subdiffusion, and thus exhibits dynamics quite similar to what was observed in the KG continuous flow model studied in Ref. \cite{Antonopoulosetal2014}.

\section{Conclusions}\label{conclusions}

In this paper we investigated the connection between regimes of ``weak'' and ``strong'' chaos  and diffusive dynamics in a multi-dimensional system of coupled symplectic McMillan maps, with the aim of comparing our results with analogous studies by S. Flach and coworkers on a Klein-Gordon Hamiltonian in the presence of disorder and nonlinearity. Our main idea was to explore whether coupled symplectic maps can be used to study wavepacket subdiffusive spreading in disordered media, thus avoiding the tedious integration of large systems of ordinary differential equations that describe Hamiltonian particle chains such as the KG model.

Our statistical approach was based on the computation of sums of position coordinates, in the spirit of the central limit theorem and nonextensive statistical mechanics, approximating the probability distribution functions by $q$-Gaussians, whose index $q>1$ is connected with weak chaos, and $q=1$ with strong chaos.

Let us summarize our findings in reference to the questions we raised in Sec. \ref{CLT_approach}: First, we demonstrated that our CMM analogue of a disordered nonlinear system of Hamiltonian oscillators displays chaotic diffusion connected with two main epochs: A relatively short one of super-diffusion followed by a second subdiffusive stage with exponents quite smaller than 1. This is in contrast to diffusive motion in a disordered KG chain where only one epoch is observed \cite{Flach2009,Antonopoulosetal2014}. Secondly, we studied CMMs with increasingly higher dimensionality to explore the relation between complex statistics and subdiffusion and verified in all cases that to observe subdiffusive motion the coupling strength has to be moderately large and independent of dimensionality. Thus, in comparison with our findings for the KG system in Ref. \cite{Antonopoulosetal2014}, the dynamics and statistics of diffusive motion in the CMM model yield analogous results. 

Finally, we showed that subdiffusion is accompanied by weak and strong chaos (with $q>1$ and $q=1$-Gaussian pdfs respectively) regardless of the dimensionality of the system, but depending on the coupling strength, in analogy to a disordered KG chain at different energy ranges. The key point behind this remark is that in the KG system nonlinearity is measured by the amplitude of the initial excitation that determines the total energy and diffusive/statistical properties of the dynamics, whereas in the coupled map model chaotic diffusion is measured by the magnitude of the coupling strength which is proportional to how much the CMMs deviate from the integrable $\epsilon=0$ limit (see Eq. \eqref{potential_subsection_Results_2012_09_07}). Finally, in the coupled map model disorder does not seem to play an equally important role as in the KG system.

Thus, even though there are substantial differences in the local dynamics of the two systems considered here, our results suggest that coupled symplectic maps may be used to model subdiffusion and weakly chaotic behavior in certain classes of multi-dimensional nonlinear oscillators. For such systems, employing discrete iterative systems can circumvent costly numerical algorithms needed to integrate the corresponding continuous time equations of motion and identify maps through complicated discretization techniques.

\section*{Acknowledgements}
One of us (T. B.) acknowledges many interesting discussions on coupled maps with Professor C. Tsallis. We are also grateful to the anonymous referees for their constructive feedback that helped us improve the manuscript and to the HPCS Laboratory of the TEI of Western Greece for providing the computer facilities where all our simulations were performed. C. G. A. was partially supported by the ``EPSRC EP/I032606/1'' grant of the University of Aberdeen. This research has been co-financed by the European Union (European Social Fund - ESF) and Greek national funds through the Operational Program ``Education and Lifelong Learning'' of the National Strategic Reference Framework (NSRF) - Research Funding Program: THALES - Investing in knowledge society through the European Social Fund. 





\providecommand{\noopsort}[1]{}\providecommand{\singleletter}[1]{#1}%

\end{document}